\documentclass[aps,prc,twocolumn,amsmath,amssymb,preprintnumbers,10pt]{revtex4-1}

\usepackage{graphicx}
\usepackage{dcolumn}
\usepackage{bm}
\usepackage{amssymb}
\usepackage{esvect}
\usepackage{xcolor}
\usepackage{braket}
\usepackage{mathtools}
\usepackage{lineno}
\usepackage{color}
\usepackage{soul}
\usepackage{comment}
\usepackage{multirow}
\usepackage[normalem]{ulem}

\newcommand{\sj}[6]{ \begin{Bmatrix}
  #1 & #2 & #3 \\
  #4 & #5 & #6 
 \end{Bmatrix}}

\def\nuc#1#2{\relax\ifmmode{}^{#1}{\protect\text{#2}}\else${}^{#1}$#2\fi}

\newcommand{\be}{\begin{eqnarray}}
\newcommand{\ee}{\end{eqnarray}}

\newcommand{\bwt}{\begin{widetext}}
\newcommand{\ewt}{\end{widetext}}

\begin{document}

\title{Eikonal calculation of $(p,3p)$ cross sections for neutron-rich nuclei}


\author{M.~G\'omez-Ramos}
\email[]{mgomez40@us.es}

\affiliation{Departamento de FAMN, Universidad de Sevilla, Apartado 1065, 41080 Sevilla, Spain.}
\affiliation{Institut f\"ur Kernphysik, Technische Universit\"at Darmstadt, D-64289 Darmstadt, Germany}


\begin{abstract}
In this work, we present the first, to our knowledge, theoretical description of two-proton removal reactions with proton target $(p,3p)$ for medium-mass nuclei at intermediate energies and present cross sections for the different bound states of the residual nucleus with two fewer protons. The description of the reaction assumes two sequential ``quasifree'' collisions between the target and removed protons and considers eikonal propagation in between. The formalism is applied to the reactions $^{12}\mathrm{C}(p,3p)^{10}\mathrm{Be}$, $^{28}\mathrm{Mg}(p,3p)^{26}\mathrm{Ne}$  and $^{54}\mathrm{Ca}(p,3p)^{52}\mathrm{Ar}$, finding reasonable agreement to experimental data for the $^{12}$C target and an overestimation of a factor $\sim3$ for the more neutron-rich and $^{54}$Ca, which is similar to the results found in two-proton knockout experiments with $^9$Be and $^{12}$C targets. 
\end{abstract}


\date{\today}%
\maketitle

\section{Introduction\label{sec:intro}}

Two-proton knockout reactions from neutron-rich nuclei using $^9$Be and $^{12}$C at intermediate energies have been shown to proceed as a direct reaction \cite{Baz03} and were shown to be able to populate very exotic nuclei via the removal of two protons from already proton-deficient nuclei \cite{Baz03,Fri05}. Equivalently, two-neutron knockout reactions from proton-rich nuclei have been used to study very neutron-rich nuclei \cite{Yon06}. The analysis of these reactions using an eikonal sudden description \cite{Tos04,Tos04b} has yielded significant results on the structure of these nuclei \cite{Fri06,Gad06,Gad07,Bas07,Adr08} and on the effect of nuclear correlations on the observables of the reactions, and therefore their value as a probe of these correlations \cite{Tos04b,Sim09,Sim09b,Yon06,wim12}. The development of new hydrogen-target detectors, such as active targets \cite{Bec15}, or MINOS \cite{Obe14}, where a thick liquid-hydrogen target is coupled to a vertex tracker for the recoil protons, has opened the use of proton-induced reactions as reliable probes to explore exotic nuclei, where the reaction mechanism can be explored and understood thanks to the tracking of the paths of the outgoing particles. Therefore, two-proton removal reactions with proton targets, or $(p,3p)$, appear as an appealing probe to produce exotic nuclei by removing two-protons from already proton-deficient species, and to be able to study their properties thanks to the simpler reaction mechanism and the possibility for proton-tracking. Unfortunately, as was the case for one-neutron removal, the reaction models used with heavier targets \cite{Han03} are not applicable for the case of the proton target, due to the significant recoil of the target proton. Models considering a ``quasi-free'' interaction between removed and target protons have been more successful in the description of the experimental data for reactions with proton targets $(p,2p)$ \cite{Jacob:1966,aum13,mor15,oga15}, so a similar approach for $(p,3p)$ reactions seems promising and is required in order to fully exploit $(p,3p)$ reactions as a spectroscopic tool using the experimental possibility that hydrogen active targets provide. This need has been indicated in previous publications \cite{Tan19} where the lack of such a theory has hindered the analysis of the experimental data.

This work aims to provide a theoretical formalism for $(p,3p)$ reactions, based on the assumption of ``quasi-free'' collision between the target and removed protons, and is structured as follows: Section~\ref{sec:theo} presents the theoretical formalism and briefly shows its derivation, section~\ref{sec:results} presents calculations of $(p,3p)$ reactions for the stable $^{12}$C target as validation of the theory and results for the neutron-rich targets $^{28}$Mg and $^{54}$Ca. Finally, section~\ref{sec:summary} presents the conclusions and summary of this work as well as future extensions.

\section{Theoretical framework\label{sec:theo}}

\begin{figure}[tb]
\begin{center}
 {\centering \resizebox*{\columnwidth}{!}{\includegraphics{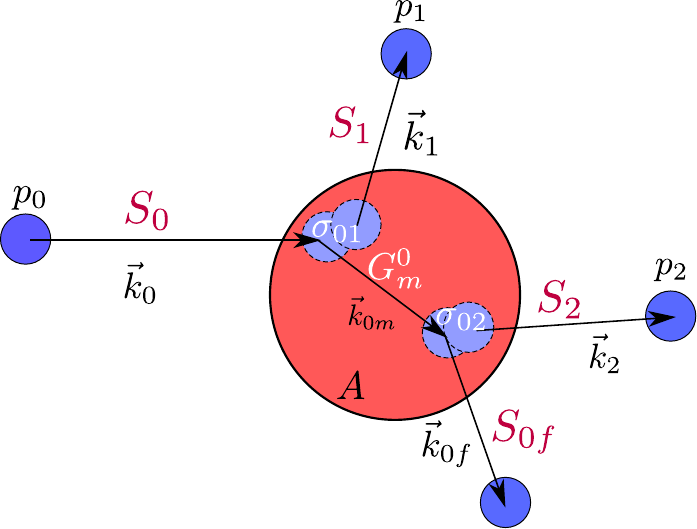}} \par}
\caption{\label{fig:draw}Schematic description of the reaction model considered in this work, two sequential collisions of the incoming proton with the removed protons. The labels used in this work are presented in black, indicating as well the wave-numbers of the protons at each step of the process. The elements appearing in the $T$-matrix for the reaction can be associated to the different steps of the trajectory of the protons and are shown close to them in purple and white.}
\end{center}
\end{figure}

We consider a process $p+(A+2) \rightarrow 3p+A $, where a projectile proton collides with the target nucleus $A+2$ removing two protons from it, with the remaining nucleus $A$ remaining bound. For the derivation, we will assume an infinite mass for $A$. Following the results from \cite{Fro20}, we describe the process for this collision as two sequential and independent collisions between the projectile proton and two protons of the target, with the residual nucleus $A$ remaining as an inert spectator. We will also assume that the reaction occurs fast enough for the internal degrees of freedom of $A$ to remain frozen during the collision, so that the removal of the two protons does not alter the state of $A$. For ease of description, in this derivation, the protons will be treated as distinguishable, since, following Goldberger and Watson \cite{Gold67}, it is sufficient to consider their antisymmetrization in the proton-proton interaction $V_{pp}$. As such, $p_0$ corresponds to the incoming proton, with momentum $\hbar\mathbf{k}_0$. $p_0$ is then assumed to collide with the first proton $p_1$, which is expelled with an asymptotic momentum $\hbar\mathbf{k}_1$. Then $p_0$ collides with a second proton $p_2$, and both escape the nucleus with momenta  $\hbar\mathbf{k}_{0f}$ and $\hbar\mathbf{k}_2$ respectively. Through momentum conservation the residual nucleus $A$ is left with momentum $\hbar\mathbf{k}_{A}=\hbar(\mathbf{k}_0-\mathbf{k}_{0f}-\mathbf{k}_1-\mathbf{k}_2)$. A diagram of the process is shown in Figure~\ref{fig:draw}. Schematically, the transition matrix for this process can be expressed as:

\begin{equation}
\begin{split}
    T\simeq \int &\mathrm{d} \xi \prod_i \left(\mathrm{d}\mathbf{r}_{i}\right)    \chi^*_{p_0}(\mathbf{r}_{0f},\mathbf{k}_{0f})\chi^*_{p_1}(\mathbf{r}_1,\mathbf{k}_{1})\chi^*_{p_2}(\mathbf{r}_2,\mathbf{k}_{2})\Phi_A(\xi_A)
    \\&[V_{02}GV_{01}](\mathbf{r}_{0f},\mathbf{r}_1,\mathbf{r}_2,\xi_A;\mathbf{r}_{0},\xi)\Phi_{(A+2)}(\xi)\chi_{p_0}(\mathbf{r}_{0},\mathbf{k}_{0}),
\end{split}
\end{equation}

where $\mathrm{d}\mathbf{r}_{i}$ denotes all radial variables involved: $\mathbf{r}_{0f},\mathbf{r}_1,\mathbf{r}_2,\mathbf{r}_0$; $\chi_{p_i}$ indicate the wavefunctions for the relative motion between the corresponding particles and residual nucleus $A$ (with infinite mass) for the corresponding asymptotic momenta, $\xi$ and $\xi_A$ correspond to the internal coordinates of $(A+2)$ and $A$ respectively, $V_{ij}$ is the interaction potential between protons $i$ and $j$ and $G$ is the propagator of the system. Due to the spectator approximation for $A$ its internal coordinates $\xi_A$ are not modified during the reaction, so neither $V$ nor $G$ depend on them. We can expand $\xi_{(A+2)}=\xi_A,\mathbf{r}_{1i},\mathbf{r}_{2i}$ , where $\mathbf{r}_{ji}$ is the position of proton $j$ ``within'' $(A+2)$. Thus, we can compute the overlap between $(A+2)$ and $A$:
\begin{equation}
    \int \mathrm{d}\xi_A \braket{A(\xi_A)|(A+2)(\xi_A,\mathbf{r}_{1i},\mathbf{r}_{2i})}=\phi_{12}(\mathbf{r}_{1i},\mathbf{r}_{2i}).
\end{equation}

We note that $\phi_{12}(\mathbf{r}_{1i},\mathbf{r}_{2i})$ is independent of the reaction and can be obtained via structure calculations such as nuclear shell model \cite{Tos04}. This results in the $T$-matrix:

\begin{equation}
\begin{split}
    T\simeq \int &\mathrm{d}\mathbf{r}_{0f}\mathrm{d}\mathbf{r}_{1}\mathrm{d}\mathbf{r}_{2}\mathrm{d}\mathbf{r}_{0}\mathrm{d}\mathbf{r}_{1i}\mathrm{d}\mathbf{r}_{2i} 
    \chi^*_{p_0}(\mathbf{r}_{0f},\mathbf{k}_{0f})\chi^*_{p_1}(\mathbf{r}_1,\mathbf{k}_{1})\\\times&\chi^*_{p_2}(\mathbf{r}_2,\mathbf{k}_{2})[V_{02}GV_{01}](\mathbf{r}_{0f}, \mathbf{r}_{1},\mathbf{r}_{2};\mathbf{r}_{0},\mathbf{r}_{1i},\mathbf{r}_{2}) \\\times&\chi_{p_0}(\mathbf{r}_{0},\mathbf{k}_{0})\phi_{12}(\mathbf{r}_{1i},\mathbf{r}_{2i}).
    \label{eq:tmat}
    \end{split}
\end{equation}
\normalsize
We now approximate propagator $G$ as \cite{Joa75}:
\begin{equation}
    G=G_{01}+GV_{01}G_{01} \simeq G_{01},
\end{equation}

which corresponds to:
\begin{equation}
\begin{split}
    &G_{01}=\frac{1}{E-H-V_{01}+i\epsilon}=\\&=\frac{1}{E-T_1-T_2-T_0-V_{02}-U_{0A}-U_{1A}-V_{2A}-V_{12}+i\epsilon},
\end{split}
\end{equation}

where the infinite mass approximation for A allows us to remove its kinetic energy. It is now our goal to reduce this four-body propagator to a one-body one. For this, we note that (since $A\rightarrow \infty$) we can group $T_2+V_{2A}\simeq H_{2A}$, and approximate for the propagator 
\begin{equation}
H_{2A}\phi_{12}(\mathbf{r}_{1i},\mathbf{r}_{2i})\sim E_{2A}\phi_{12}(\mathbf{r}_{1i},\mathbf{r}_{2i})\sim \dfrac{S_{2p}}{2}\phi_{12}(\mathbf{r}_{1i},\mathbf{r}_{2i}).
\end{equation}

This assumption should be reasonable for an energetic beam, as $p_2$ should remain frozen before its collision and $E_{2A}<<E$. We can also group $T_1+U_{1A}+V_{12}=H_{1(A+1)}$, which can be approximated as  
\begin{equation}
\chi^*_{p_1}(\mathbf{r}_1,\mathbf{k}_{1})H_{1(A+1)}=\chi^*_{p_1}(\mathbf{r}_1,\mathbf{k}_{1})E_1,
\end{equation}

and finally $U_{0A}+V_{02}\simeq U_{0(A+1)}$. With these approximations, the propagator reduces to:
\begin{equation}
\begin{split}
    G_{01}&=\frac{1}{E-E_1-E_{2A}-T_0-U_{0(A+1)}+i\epsilon}=\\&=\frac{1}{E_{0m}-T_0-U_{0(A+1)}+i\epsilon},
    \end{split}
\end{equation}

where $E_{0m}=E-E_1-E_{2A}$ and which can be interpreted as a one-body propagator for $p_0$ between collisions. We will assume the beam energy is high enough that $E_1$ is high enough that the effect of the potential can be ignored, so that $E_1=\sqrt{m_p^2+\hbar^2k_1^2}$ $(c=1)$.


We next consider the proton-proton interaction to be of zero range so that 

\begin{equation}
    V_{01}(\mathbf{r}'_0,\mathbf{r}'_1;\mathbf{r}_0,\mathbf{r}_1)=\tilde{V}_{01} \delta(\mathbf{r}'_0,\mathbf{r}'_1,\mathbf{r}_0,\mathbf{r}_1).
\end{equation}

With these approximations, the integral in Eq.~\ref{eq:tmat} is reduced to only two radial coordinates $\mathbf{r}_1, \mathbf{r}_2$, which can be interpreted as the location of the collisions between $p_0$ and $p_1$ and between $p_0$ and $p_2$ respectively:
\begin{equation}
\begin{split}
    T\simeq& \int \mathrm{d}\mathbf{r}_{1} \mathrm{d}\mathbf{r}_{2}\chi^*_{p_0}(\mathbf{r}_{2},\mathbf{k}_{0f})\chi^*_{p_1}(\mathbf{r}_1,\mathbf{k}_{1})\chi^*_{p_2}(\mathbf{r}_2,\mathbf{k}_{2})
    \\&\tilde{V}_{02}G_{0m}(E_{0m},\mathbf{r}_{2};\mathbf{r}_{1})\tilde{V}_{01}\chi_{p_0}(\mathbf{r}_{1},\mathbf{k}_{0})\phi_{12}(\mathbf{r}_{1},\mathbf{r}_{2}).
\end{split}
\end{equation}

We now introduce eikonal expressions for the wavefunctions

\begin{align}
    \chi_{p_n}(\mathbf{r}_{j},\mathbf{k}_{i})=&\dfrac{1}{(2\pi)^{3/2}}e^{-\frac{i}{\hbar v_i}\int^{z_j}_{-\infty} U(\mathbf{b}_j,z) \mathrm{d}z}e^{i\mathbf{k}_0\mathbf{r}_1}\nonumber\\=&\dfrac{1}{(2\pi)^{3/2}}S_{p0}(\mathbf{r}_1,\mathbf{k}_0)e^{i\mathbf{k}_0\mathbf{r}_1} \label{eq:smat}
\end{align}

where $z$ follows the direction of momentum $\mathbf{k}_i$, $\mathbf{b}$ is the associated impact parameter and $U$ is the optical potential between proton and core $A$ (As $A\rightarrow \infty$ we assume $U_{pA}\sim U_{p(A+1)}\sim U_{p(A+2)}$). 
With this approximation we obtain for the matrix element:

\begin{equation}
\begin{split}
T&=\dfrac{1}{(2\pi)^{6}}\int \mathrm{d}\mathbf{k}_{0m}\mathrm{d}\mathbf{r}_1\mathrm{d}\mathbf{r}_2S^*(\mathbf{r}_{2},\mathbf{k}_{0f}) S^*(\mathbf{r}_{2},\mathbf{k}_{2})\\&\tilde{V_{02}}G_{0m}(E_{0m},\mathbf{r}_{2};\mathbf{r}_{1})\tilde{V_{01}}S^*(\mathbf{r}_{1},\mathbf{k}_{1})S(\mathbf{r}_{1},\mathbf{k}_{0})\phi_{12}(\mathbf{r}_{1},\mathbf{r}_{2})\\&
e^{i(\mathbf{k}_0-\mathbf{k}_1)\mathbf{r}_1}e^{-i(\mathbf{k}_{0f}+\mathbf{k}_2)\mathbf{r}_2}.
\label{eq:t_end}
\end{split}
\end{equation}

Next let us present the formula for the total cross section, where momentum conservation has already been considered to nullify the integral over $\mathbf{k_{A}}$:

\begin{equation}
\sigma=\dfrac{1}{2\hat{J_i}^2}\sum_{S,s_0,s_{0f}}\int \mathrm{d}\mathbf{k}_1 \mathrm{d}\mathbf{k}_{0f} \mathrm{d}\mathbf{k}_{2} \delta(E_f-E_i) \dfrac{(2\pi)^4}{\hbar v_0} \left|T\right|^2, \label{eq:sig_begin}
\end{equation}

where $J_i$ is the angular momentum of $(A+2)$ $\hat{J_i}=\sqrt{2J_i+1}$, the sum over $S$ corresponds to the sum over the spin projections of $(A+2),A,p_1,p_2$ and the sums for the final and initial spin of proton $p_0$ are left explicit. We will now express $\mathrm{d}\mathbf{k_{0f}}$ in spherical coordinates and will integrate over $k_{0f}$ using $\delta(E_f-E_i)$. Thus we obtain
\begin{equation}
\begin{split}
&\int \mathrm{d}k_{0f}\delta(E_f-E_i)\mathcal{F}(k_{0f})=\mathcal{F}(\bar{k}_{0f})\left.\dfrac{1}{\dfrac{\partial E_f}{\partial k_{0f}}}\right|_{\bar{k}_{0f}}=
\\&=\dfrac{1}{\dfrac{\hbar^2 c^2\bar{k}_{0f}}{\bar{\epsilon}_{0f}}+\dfrac{\hbar^2 c^2\bar{k}_{0f}}{\bar{\epsilon}_{2}}-\dfrac{\hbar^2 c^2\bar{\mathbf{k}}_{0f}(\mathbf{k}_0-\mathbf{k}_1-\mathbf{k}_A)}{{k_{0f}}\bar{\epsilon}_{2}}}\mathcal{F}(\bar{k}_{0f})\equiv\\&\equiv j(\hat{\mathbf{k}}_{0f},\mathbf{k}_1,\mathbf{k}_A)\mathcal{F}(\bar{k}_{0f})
\end{split}
\end{equation}

where $j(\hat{\mathbf{k}}_{0f},\mathbf{k}_1,\mathbf{k}_A)$ corresponds to these kinematic factors and $\mathcal{F}(k_{0f})$ is meant to denote the rest of integral in Eq.~\ref{eq:sig_begin} and where we have applied that $k_2^2=k_{0f}^2-2\mathbf{k}_{0f}(\mathbf{k}_0-\mathbf{k}_1-\mathbf{k}_A)+|\mathbf{k}_0-\mathbf{k}_1-\mathbf{k}_A|^2$ and $\bar{\mathbf{k}}_{0f}, \bar{\epsilon}_{0f}, \bar{\epsilon}_{2}$ correspond to the momentum for $p_{0f}$ and energies of $p_{0f}$ and $p_2$ given by energy conservation.

To proceed with the derivation, we will now introduce the strong approximation that all factors in Eqs~(\ref{eq:t_end}) and (\ref{eq:sig_begin}) are slowly varying functions of the momentum compared to the exponentials in Eq.~\ref{eq:t_end} and can be replaced by a suitable average value which may depend on $\mathbf{r_1}, \mathbf{r_2}$:

\begin{equation}
    \begin{split}
        S(\mathbf{r}_{i},\mathbf{k}_{j})&\sim S(\hat{\mathbf{k}}_{0f},\mathbf{r}_{1},\mathbf{r}_{2})\\
        j(\hat{\mathbf{k}}_{0f},\mathbf{k}_1,\mathbf{k}_A)&\sim j(\hat{\mathbf{k}}_{0f},\mathbf{r}_1,\mathbf{r}_2),
    \end{split}
\end{equation}

which is a general expression, although some of the $S$-matrices may not depend on all variables. With this approximation, Eq.~\ref{eq:t_end} can be expressed as:
\begin{equation}
\begin{split}
T&\simeq\dfrac{1}{(2\pi)^{6}}\int \mathrm{d}\mathbf{r}_1\mathrm{d}\mathbf{r}_2\tilde{S_{0f}}^*(\hat{\mathbf{k}}_{0f},\mathbf{r}_{2},\mathbf{r}_{1}) \tilde{S_2}^*(\hat{\mathbf{k}}_{0f},\mathbf{r}_{2},\mathbf{r}_{1})\\\times&\tilde{G}_{0m}(E_{0m},\mathbf{r}_{2};\mathbf{r}_{1})\tilde{S_1}^*(\mathbf{r}_{1},\mathbf{r}_{2})\tilde{S_0}(\mathbf{r}_{1})|\phi_{12}|^2(\mathbf{r}_{1},\mathbf{r}_{2})\\&\tilde{V_{01}}(\mathbf{r}_{1};\mathbf{r}_{2})
\tilde{V_{02}}(\mathbf{r}_1;\mathbf{r}_{2})e^{i(\mathbf{k}_0-\mathbf{k}_1)\mathbf{r}_1}e^{-i({\mathbf{k}_{0f}+\mathbf{k}_2)\mathbf{r}_2}},
\end{split}
\end{equation}

and now we can perform the integrals over $\mathbf{k}_1$ and $\mathbf{k}_A$, noting that $T$ and $T^*$ would produce equivalent and complex conjugate exponentials with spatial variables $(\mathbf{r}_1,\mathbf{r}_2)$ and $(\mathbf{r}'_1,\mathbf{r}'_2)$:
\begin{equation}
\begin{split}
    \int &\mathrm{d}\mathbf{k}_1  \mathrm{d}\mathbf{k}_{2}\left[e^{-i\mathbf{k}_1(\mathbf{r}_1-\mathbf{r}'_1)}e^{i\mathbf{k}_2(\mathbf{r}_2-\mathbf{r}'_2)}\right] =\\&=(2\pi)^6 \delta(\mathbf{r}_2-\mathbf{r}_2') \delta(\mathbf{r}_1-\mathbf{r}'_1),
\end{split}
\end{equation}

which produces the nice result that the integral reduces to an incoherent sum over the two collision points $\mathbf{r}_1$ and $\mathbf{r}_2$, which is consistent with the semiclassical assumption.

\begin{equation}
\begin{split}
\sigma&=\dfrac{1}{2\hat{J}_i^2}\sum_{S,s_0,s_{0f}}\int  \mathrm{d}\hat{\mathbf{k}}_{0f}  \mathrm{d}\mathbf{r}_1 \mathrm{d}\mathbf{r}_2 \\\times&\dfrac{1}{\hbar v_0 (2\pi)^2} j(\hat{\mathbf{k}}_{0f},\mathbf{r}_1,\mathbf{r}_2)|\tilde{S_{0f}}|^2(\hat{\mathbf{k}}_{0f},\mathbf{r}_{2},\mathbf{r}_{1}) |\tilde{S_2}|^2(\hat{\mathbf{k}}_{0f},\mathbf{r}_{2},\mathbf{r}_{1})\\\times&\tilde{V_{02}}(\mathbf{r}_1;\mathbf{r}_{2})|^2|\tilde{G}_{0m}|^2 (E_{0m},\mathbf{r}_{2};\mathbf{r}_{1})\\\times&|\tilde{S_1}|^2(\mathbf{r}_{1},\mathbf{r}_{2})|\tilde{V_{01}}(\mathbf{r}_{1};\mathbf{r}_{2})|^2|\tilde{S_0}(\mathbf{r}_{1})|^2|\phi_{12}|^2(\mathbf{r}_{1},\mathbf{r}_{2}).
\end{split}
\end{equation}

We will now use the eikonal expression for the propagator \cite{Kaw92,Gold67}:
\begin{align}
\begin{split}
 G_{0m}(E_{0m},\mathbf{r}_{2};\mathbf{r}_{1})=&-\dfrac{\epsilon_{0m}}{2\pi \hbar^2c^2} \dfrac{e^{-\frac{i}{\hbar v_{0m}}\int_{z_1}^{z_2}U(\mathbf{b},z) \mathrm{d}z}}{|\mathbf{r_2-r_1}|}\\\times &e^{-ik_{0m}|\mathbf{r_2-\mathbf{r_1}}|} \label{eq:propag},
 \end{split}
\end{align}

where we assume the energy $E_{0m}$ is large enough that the integral can be performed along the straight line between $\mathrm{r_2}$ and $\mathrm{r_1}$ and $\hbar k_{0m}$ is the momentum associated to $E_{0m}$

We note that a similar method for calculating the $(p,2p)$ total cross section was presented in \cite{aum13}. We must now choose the average values to be used for each quantity. For the S-matrices and propagator $G_{0m}$ Eqs.~(\ref{eq:smat}) and (\ref{eq:propag}) provide eikonal expressions. For $|\tilde{V_{01}}(\mathbf{r}_{1};\mathbf{r}_{2})|^2$ we will assume on-shell quasifree collisions between the protons and also assume that the removed protons are initially at rest, although their binding energy, taken as $S_{2p}/2$, will be taken into account. Therefore in the first collision,  proton $p_0$ collides with momentum $\hbar \mathbf{k}_0$ with proton $p_1$ and both are emitted with momenta $\hbar \mathbf{k}_{0m}$ and $\hbar \mathbf{k}_1$ respectively. We note that the direction of $\mathbf{k}_{0m}$ is parallel to $(\mathbf{r}_2-\mathbf{r}_1)$, which is consistent with eikonal propagation. As such, positions $\mathbf{r}_1$ and $\mathbf{r}_2$ along with energy and momentum conservation in the quasi-free collision define $k_{0m}$ and  $\mathbf{k}_1$. For the second collision, assuming the proton impacts with momentum $\hbar \mathbf{k}_{0m}$, the direction of $\mathbf{k}_{0f}$ restricts the value of $\mathbf{k_2}$ and the modulus of $k_{0f}$, to be used in $S_2$ and $S_{0f}$, taking $\mathbf{k}_A$ as 0. Finally, $|\tilde{V_{01}}(\mathbf{r}_{1};\mathbf{r}_{2})|^2$ can be obtained from the free $p-p$ cross section (omitting Coulomb interaction) assuming the Born 
approximation:
\begin{equation}
|\tilde{V_{01}}(\mathbf{r_1};\mathbf{r_{2}})|^2\simeq|\tilde{T_{01}}(\mathbf{r_1};\mathbf{r_{2}})|^2.
\end{equation}

Now we move to the NN center of mass system, introducing the adequate M\"oller factor

\begin{equation}
|\tilde{T_{01}}(\mathbf{r}_1;\mathbf{r}_{2})|^2=|\tilde{T_{01}}(\mathbf{r}_1;\mathbf{r}_{2})|_{NN}^2 \dfrac{\epsilon_{cmi,0}\epsilon_{cmi,1}\epsilon^{2}_{cmf,1}}{(m_Nc^2-S_{2p}/2)\epsilon_{0}\epsilon_1\epsilon_{0m}},
\end{equation}

where, since the collision happens between a proton and a pseudoproton of mass $(m_Nc^2-S_{2p}/2)$, the center of mass energies of both particles are not the same before the collision ($\epsilon_{cmi,0}$ being the one of the proton and $\epsilon_{cmi,1}$ the one of the pseudoproton) but they are the same after the collision $\epsilon_{cmf}$, since both outgoing particles are protons. The relation between $\epsilon_{cmi}$ and $\epsilon_{cmf}$ can be obtained through the Mandelstan variable $s$. We can relate the NN center of mass T matrix to the angular differential free NN cross section \cite{oga15}, which we will approximate by the total cross section divided by $4\pi$:

\begin{equation}
|\tilde{T_{01}}(\mathbf{r}_1;\mathbf{r}_{2})|_{NN}^2=\dfrac{4(2\pi)^2\hbar^4c^4}{\epsilon_{cmf,1}^2}\dfrac{\mathrm{d}\sigma}{\mathrm{d}\Omega}=2\dfrac{4(2\pi)^2\hbar^4c^4}{\epsilon_{cmf,1}^2}\dfrac{\sigma_{NN}}{4\pi}
\end{equation}

 (factor 2 due to antisymmetrization) and we get:

\begin{equation}
\begin{split}
&|\tilde{V_{01}}(\mathbf{r}_1;\mathbf{r}_{2})|^2\simeq 4(2\pi)\hbar^4c^4\sigma_{NN}(\mathbf{r}_1;\mathbf{r}_2)\\\times&\dfrac{\epsilon_{cmi,0}\epsilon_{cmi,1}}{(m_Nc^2-S_{2p}/2)\epsilon_{0}\epsilon_1\epsilon_{0m}}\equiv\sigma_{NN}(\mathbf{r}_1;\mathbf{r}_2)f(\mathbf{r}_1,\mathbf{r}_2),
\end{split}
\end{equation}

where the factors multiplying the nucleon-nucleon cross section $\sigma_{NN}$ are included in $f(\mathbf{r}_1,\mathbf{r}_2)$. We note that this approximation for the nucleon-nucleon interaction makes it spin-independent. Therefore the projection of the spin of $p_0$ must be the same in the incoming and outgoing channels, so that $\sum_{s_0,s_{0f}}=2$, which cancels with the factor 1/2 in Eq.~(\ref{eq:sig_begin}). This results in the following expression for the cross section:

\begin{equation}
\begin{split}
\sigma&=\dfrac{1}{\hat{J_i^2}}\sum_{S}\int  \mathrm{d}\hat{\mathbf{k}}_{0f}  \mathrm{d}\mathbf{r}_1 \mathrm{d}\mathbf{r}_2 \\\times&\dfrac{1}{\hbar v_0 (2\pi)^2} j(\hat{\mathbf{k}}_{0f},\mathbf{r}_1,\mathbf{r}_2)f_{02}(\hat{\mathbf{k}}_{0f},\mathbf{r}_1,\mathbf{r}_2)f_{01}(\mathbf{r}_1,\mathbf{r}_2) \\\times&|\tilde{S_{0f}}|^2(\hat{\mathbf{k}}_{0f},\mathbf{r}_{2},\mathbf{r}_1)|\tilde{S_2}|^2(\hat{\mathbf{k}}_{0f},\mathbf{r}_{2},\mathbf{r}_{1})\sigma_{02}(\mathbf{r}_1;\mathbf{r}_2)\tilde{G}^m_0(\mathbf{r}_{2},\mathbf{r}_{1})|^2\\\times&|\tilde{S_1}|^2(\mathbf{r}_{1},\mathbf{r}_{2})\sigma_{01}(\mathbf{r}_1;\mathbf{r}_2)|\tilde{S_0}(\mathbf{r}_{1})|^2|\phi_{12}|^2(\mathbf{r}_{1},\mathbf{r}_{2}),
\end{split}
\end{equation}

which allows for a very classical interpretation: proton $p_0$ penetrates inside the nucleus with probability $|\tilde{S}_0|^2$ until it collides with cross section $\sigma_{01}$ with one of the protons $p_1$ in the nucleus at $\mathbf{r}_1$ and then propagates to $\mathbf{r}_2$ through $|\tilde{G}_{0m}|^2$ where it collides with the second proton $p_2$ with cross section $\sigma_{02}$, with the probability of finding both protons at these positions given by $|\phi_{12}|^2$ for a certain state of the residual core. Afterwards, all three protons must escape the nucleus with probabilities $|\tilde{S}_{0f}|^2,|\tilde{S}_{1}|^2$ and $|\tilde{S}_{2}|^2$, for them to be detected. See Figure~\ref{fig:draw} for a schematic with the factors associated to the paths of the protons. For a more compact expression, which is also more comparable with the ones in the literature, we will include the spin average and sum in $|\phi_{12}|^2$, noting that it is the only term which is sensitive to the spins of the protons:

\begin{equation}|\tilde{\phi}_{12}|^2(\mathbf{r}_{1},\mathbf{r}_{2})=\dfrac{1}{\hat{J_i}^2}\sum_{S}|\phi_{12}|^2(\mathbf{r}_{1},\mathbf{r}_{2})
\end{equation}
resulting in:
\begin{equation}
\begin{split}
\sigma&=\int  \mathrm{d}\hat{\mathbf{k}}_{0f}  \mathrm{d}\mathbf{r}_1 \mathrm{d}\mathbf{r}_2 \dfrac{1}{\hbar v_0 (2\pi)^2} j(\hat{\mathbf{k}}_{0f},\mathbf{r}_1,\mathbf{r}_2)f_{02}(\hat{\mathbf{k}}_{0f},\mathbf{r}_1,\mathbf{r}_2) \\\times&f_{01}(\mathbf{r}_1,\mathbf{r}_2)|\tilde{S_{0f}}|^2(\hat{\mathbf{k}}_{0f},\mathbf{r}_{2},\mathbf{r}_1)|\tilde{S_2}|^2(\hat{\mathbf{k}}_{0f},\mathbf{r}_{2},\mathbf{r}_{1})\sigma_{02}(\mathbf{r}_1;\mathbf{r}_2)\\\times&\tilde{G}^m_0(\mathbf{r}_{2},\mathbf{r}_{1})|^2|\tilde{S_1}|^2(\mathbf{r}_{1},\mathbf{r}_{2})\sigma_{01}(\mathbf{r}_1;\mathbf{r}_2)|\tilde{S_0}(\mathbf{r}_{1})|^2|\tilde{\phi}_{12}|^2(\mathbf{r}_{1},\mathbf{r}_{2}).
\end{split}
\end{equation}


\subsection{Overlap function}

We present briefly the calculation of the square of the overlap function $|\phi_{12}|^2$ from nuclear structure inputs: two-nucleon amplitudes (TNA) and single-particle wavefunctions. We start from the expression from \cite{Sim09}:
\begin{equation}
\begin{split}
    &|\phi_{12}(\mathbf{r}_1,\mathbf{r}_2)|^2=\\&\dfrac{1}{J_i^2}\sum_{M_iM_f} \left\langle\Psi^{(F)}_{J_iM_i}|\Psi^{(F)}_{J_iM_i}\right\rangle=\sum_{\alpha\alpha'I}\dfrac{C_\alpha^{J_iJ_fI}C_{\alpha'}^{J_iJ_fI}D_\alpha D_{\alpha'}}{I^2}\\\times&\sum_{m_1m_2m_1'm_2'} <j_1m_1j_2m_2|I\mu><j_1'm_1'j_2'm_2'|I\mu>\\\times&[(\phi^{m_1'}_{j_1'}|\phi^{m_1}_{j_1})(\phi^{m_2'}_{j_2'}|\phi^{m_2}_{j_2})+(\phi^{m_2'}_{j_2'}|\phi^{m_2}_{j_2})(\phi^{m_1'}_{j_1'}|\phi^{m_1}_{j_1})\\&-(\phi^{m_1'}_{j_1'}|\phi^{m_2}_{j_2})(\phi^{m_2'}_{j_2'}|\phi^{m_1}_{j_1})-(\phi^{m_2'}_{j_2'}|\phi^{m_1}_{j_1})(\phi^{m_1'}_{j_1'}|\phi^{m_2}_{j_2})],
    \label{eq:overlap}
\end{split}
\end{equation}

where $\Psi^{(F)}_{J_iM_i}$ is the overlap function $\braket{A|A+2}$ with $J_i,M_i$ and $J_f,M_f$ being the total and magnetic angular momenta of $A+2$ and $A$ respectively. $C_\alpha^{J_iJ_fI}$ corresponds to the TNA, $\alpha=l_1,j_1,l_2,j_2$, the orbital and angular momenta of the two removed protons, $m_i$ the corresponding magnetic angular momentum, $D_\alpha=1/\sqrt{2(1+\delta_{\alpha\alpha'})}$, $\mathbf{I}=\mathbf{j_1}+\mathbf{j_2}$ and the bracket $(\phi^{m_2'}_{j_2'}|\phi^{m_2}_{j_2})$ corresponds to:
\begin{equation}
\begin{split}
(\phi^{m_1'}_{j_1'}|\phi^{m_1}_{j_1})=&\sum_{\lambda_1\lambda_1'\sigma} <l_1\lambda_1s\sigma|j_1m_1><l_1'\lambda_1's\sigma|j_1'm_1'>\\& u_{l_1j_1}(r_1)Y_{l_1\lambda_1}(\hat{\mathbf{r_1}})u^*_{l_1'j_1'}(r_1)Y^*_{l_1'\lambda_1'}(\hat{\mathbf{r_1}}),
\end{split}
\end{equation}

where $ u_{lj}$ is the radial wavefunction, evaluated at $S_{2p}/2$ \cite{Sim09}, $Y_{l\lambda}$ the spherical harmonic with magnetic angular momentum $\lambda$ and $s$ and $\sigma$ are the spin of the proton and its projection. Note that in Eq.~(\ref{eq:overlap}), the first bracket in each term is evaluated for $\mathbf{r}_1$ while the second bracket is evaluated for $\mathbf{r}_2$.

After some angular momentum algebra, we reach the following expression:

\begin{equation}
\begin{split}
&|\tilde{\phi}_{12}|^2(\mathbf{r_1},\mathbf{r_2})=\sum_{\alpha\alpha'I}C_\alpha^{J_iJ_fI}C_{\alpha'}^{J_iJ_fI}D_\alpha D_{\alpha'}\dfrac{\hat{l_1}\hat{l_1'}\hat{l_2}\hat{l_2'}\hat{j_1}\hat{j_2}\hat{j_1'}\hat{j_2'}}{16\pi^2}
\\\times &(-)^{-s_1-s_2+j_2+j_2'}\sum_{L}P^L(\mathbf{r_1}\cdot\mathbf{r_2})\\\times&\left[\left((-)^{I-2j_2'}<l_10,l_1'0|L0><l_20,l_2'0|L0>\sj{l_1}{s_1}{j_1}{j_1'}{L}{l_1'}
\right.\right.\\\times&\left.\sj{l_2}{s_2}{j_2}{j_2'}{L}{l_2'}\sj{j_1}{I}{j_2}{j_2'}{L}{j_1'}\right)\\\times&(u_{l_1j_1}(r_1)u^*_{l_1'j_1'}(r_1)u_{l_2j_2}(r_2)u^*_{l_2'j_2'}(r_2))+\\\times&+u_{l_2j_2}(r_1)u^*_{l_2'j_2'}(r_1)u_{l_1j_1}(r_2)u^*_{l_1'j_1'}(r_2))-\\-&\left(<l_10,l_2'0|L0><l_20,l_1'0|L0>\sj{l_1}{s_1}{j_1}{j_2'}{L}{l_2'}\sj{l_2}{s_2}{j_2}{j_1'}{L}{l_1'}\right.\\\times&\left.\sj{j_1}{I}{j_2}{j_1'}{L}{j_2'}\right)(u_{l_2j_2}(r_1)u^*_{l_1'j_1'}(r_1)u_{l_1j_1}(r_2)u^*_{l_2'j_2'}(r_2))+\\\times&+\left.u_{l_1j_1}(r_1)u^*_{l_2'j_2'}(r_1)u_{l_2j_2}(r_2)u^*_{l_1'j_1'}(r_2))\right],
\end{split}
\end{equation}

where we note that the angular dependence on $\hat{\mathbf{r}}_1$ and $\hat{\mathbf{r}}_2$ is reduced to the Legendre polynomial $P^L(\mathbf{r_1}\cdot\mathbf{r_2})$, which allows for more efficient treatment of the overlap wavefunction, which can be stored as a function of $r_1$, $r_2$ (only the modulus) and $L$, instead of $\mathbf{r}_1$ and $\mathbf{r}_2$.

\section{Numerical results\label{sec:results}}

For all the following calculations the optical potential between proton and nucleus has been taken from the global Dirac parametrization \cite{Ham90, Coo93}, while the proton-proton elastic cross section has been taken from the parametrization by Bertulani and De Conti \cite{Ber10}. The single-particle wavefunctions have been computed using Woods-Saxon potentials with diffusivity $a=0.7$ fm and a radius adjusted to reproduce the rms radius from Hartree-Fock calculations using the SkX interaction \cite{Bro98}. The integration over $r_1$ and $r_2$ is extended up to the radius where the single-particle wavefunctions is reduced to $1/1000$ of their maximum value (the maximum radius among the wavefunctions) and the step of integration is taken as 0.2 fm, which was found to give converged results to $\sim$5\%.

\subsection{$^{12}\mathrm{C}(p,3p)^{10}\mathrm{Be}$\label{subsec:cbe}}

We first present a study of the $^{12}$C$(p,3p)^{10}$Be reaction. Admittedly, the $^{12}$C nucleus is rather light while the formalism developed in this work is meant for heavier systems. An estimation of the effect of the finite mass of the nucleus can be obtained from \cite{Kaw92}, which presents a very similar result to ours in the context of the study of the total nucleon-nucleus cross section. In \cite{Kaw92}, the cross section is scaled by a factor of $\left(\frac{A}{A+1}\right)^4$, which in our case would introduce a factor of 0.7 in the cross sections. We believe the ambiguities in the optical potentials may introduce larger uncertainties in the total cross section, so we will not consider this factor in the following. The nuclear structure of $^{12}$C is well known and there exist experimental data for fragmentation of $^{12}$C on proton targets to produce $^{10}$Be \cite{Ols83}, so we consider this reaction a reasonable one to benchmark our calculations.

The TNA for the calculation have been obtained using the WBT interaction \cite{WBT} in the shell model code \textsc{oxbash} \cite{oxbash} (calculations using the interaction by Cohen and Kurath \cite{Coh65} show consistent TNA). Those TNA deemed too small have been excluded from the calculation. In Table~\ref{tab:12CTNA} the used TNA are presented. In order to compare with the total experimental cross section in \cite{Ols83}, the calculation has been performed at 1.05 GeV/A. Only states under the neutron emission threshold for $^{10}$Be have been considered and their energies have been taken as the experimental ones.

\begin{table*}
    \centering
    \begin{tabular}{cc|ccccccccc}
    $J^\pi_f$&$E_x$ (MeV)& $[1s_{1/2}]^2$& $[1p_{3/2}]^2$ & $[1p_{1/2}1p_{3/2}]$ & $[1p_{1/2}]^2$ &  $[1d_{5/2}]^2$ &  $[1d_{3/2}]^2$ &  $[2s_{1/2}]^2$ &  $[2s_{1/2}1d_{5/2}]$ & $[2s_{1/2}1d_{3/2}]$
         \\ \hline
      $0^+$& 0.00& ---& 1.461 & --- &0.706 &  -0.064 &  -0.056 &  -0.056 &  --- & ---   \\
      $2^+$&3.37& ---& 2.060 & -0.854 & --- &  0.038 & 0.020 &  --- &  --- & ---\\
       $2^+$&5.96& ---&-0.419 & 1.204 & --- & 0.028 &  0.016 &  --- &  -0.020 & -0.010\\
       $0^+$&6.18& -0.018& -0.165 & --- & 0.145 &  --- &  0.016 &  --- &  --- & --- \\
       \hline\hline
       $J^\pi_f$&$E_x$ (MeV)& $[1p_{3/2}1s_{1/2}]$& $[1p_{1/2}1s_{1/2}]$ & $[1d_{5/2}1p_{3/2}]$ & $[1d_{5/2}1p_{1/2}]$&  $[1d_{3/2}1p_{3/2}]$ &  $[1d_{3/2}1p_{1/2}]$ & $[2s_{1/2}1p_{3/2}]$ &  $[2s_{1/2}1p_{1/2}]$ &   \\\hline
       $1^-$&5.96&-0.247 & 0.135 & -0.117& ---&  0.069 & -0.077 & 0.065 &  0.051 & \\
       $2^-$& 6.26& 0.045& --- & -0.048 & -0.055&  0.105 & -0.021 & --- &  --- &  
       
    \end{tabular}
    \caption{TNA used for the $^{12}$C$(p,3p)^{10}$Be reaction, not including the isospin Clebsch-Gordan factor, obtained using the WBT interaction.}
    \label{tab:12CTNA}
\end{table*}

The results are presented in Table.~\ref{tab:12Cxs}. We must note that since the kinematics of the removed protons was not measured in the experiment, another process that could result in $^{10}$Be is a $(p,2p)$ reaction producing an excited $^{11}\mathrm{B}^*$ which then decays by emission of a proton and thus competes with the process studied here \cite{Ols83}. In a recent measurement of the $^{12}$C$(p,2p)$ reaction at 398 MeV/A, the cross section for the $^{12}$C$(p,2p)^{11}$B process was found to be 18.1(2.0) mb while the cross section for the decay of $^{11}$B$^*$ into $^{10}$Be (distinguished by imposing a coincidence of a forward proton and $^{10}$Be) was 0.8(0.3) \cite{Panthesis}. Assuming the same proportion for the reaction at 1.05 GeV/A, given that the $^{12}$C$(p,2p)^{11}$B cross section was of 30.9(3.4) mb, this would yield a cross section for excitation-evaporation of 1.4(0.6) mb, leaving 2.0(0.7) mb for the direct removal of the two protons studied in this work, which gives reasonable agreement with the result of our calculations, and thus we consider this result a validation of this work. It should be remarked that in \cite{Tos04b,Sim11}, for two-proton removal from $^{12}$C using a carbon target at the same beam energy, the excitation-evaporation process was not removed and good agreement between theory and experiment was found still, which could indicate differences between proton and carbon targets when populating proton-unbound excited states of $^{11}$B.

\begin{table}
    \centering
    \begin{tabular}{cc|c}
     $J^\pi_f$&$E_x$ (MeV)& $\sigma$ (mb)
         \\ \hline
         $0^+$&0.00&0.86\\
         $2^+$&3.37&0.68\\
         $2^+$&5.96&0.27\\
         $1^-$&5.96&$9.2\cdot10^{-3}$\\
         $0^+$&6.18&$4.1\cdot10^{-3}$\\
         $2^-$& 6.26&$5.5\cdot10^{-4}$\\
         \hline
         Total&&1.82\\
         Exp.\cite{Ols83}&&3.41(0.54)\\
         Exp. (direct $(p,3p)$)&&2.0(0.7)*
    \end{tabular}
    \caption{Cross sections for the $^{12}$C$(p,3p)^{10}$Be$(J^\pi_f)$ reaction at 1.05 GeV/A\newline $^*$See text}
    \label{tab:12Cxs}
\end{table}

\subsection{$^{28}\mathrm{Mg}(p,3p)^{26}\mathrm{Ne}$\label{subsec:mgne}}

In this section, we explore the $^{28}\mathrm{Mg}(p,3p)^{26}\mathrm{Ne}$ reaction at 250 MeV/A. Although no experimental data exists for this reaction, the equivalent two-nucleon knockout reaction with beryllium target has been measured \cite{Baz03} and thoroughly studied \cite{Tos04,Tos04b,Sim10}, so we find this reaction is a good benchmark to compare the $(p,3p)$ and two-proton knockout reactions when populating different excited states. The structure inputs are the same as those from \cite{Tos04}, the TNA are presented in Table~\ref{tab:28MgTNA} for completeness.
\begin{table*}
    \centering
    \begin{tabular}{cc|cccccc}
    $J^\pi_f$&$E_x$ (MeV)& $[1d_{3/2}]^2$&  $[1d_{3/2}1d_{5/2}]$ & $[1d_{5/2}]^2$ & $[2s_{1/2}1d_{3/2}]$ & $[2s_{1/2}1d_{5/2}]$ & $[2s_{1/2}]^2$
         \\ \hline
      $0^+$& 0.00& -0.301& ---& -1.047&--- &  --- &  -0.305 \\
      $2^+$&2.02& -0.050& 0.374 & -0.637 & -0.061 &  -0.139 &--- \\
       $4^+$&3.50& ---&0.331 & 1.596 & --- & --- &  ---\\
       $2^+$&3.70& 0.047& -0.072 & 0.853 & 0.161 & 0.176&---\\
       
    \end{tabular}
    \caption{TNA used for the $^{28}$Mg$(p,3p)^{26}$Ne reaction, taken from \cite{Tos04}.}
    \label{tab:28MgTNA}
\end{table*}

We are interested in whether $(p,3p)$ and two-proton knockout reactions populate equally the excited states of the residual nucleus, as such we present in Table~\ref{tab:28Mgxs} the cross sections for the four considered states for the $(p,3p)$ reaction at 250 MeV/A, a typical energy for radioactive beam facilities, and the two-proton knockout reaction at 82.3 MeV/A \cite{Baz03}, as well as their ratio to the total cross section. Experimental results for the two-proton knockout reaction are presented as well. The theoretical and experimental results for the two-proton knockout reaction are taken from \cite{Sim10} and \cite{Tos04} respectively.

In general, we find the distribution of the cross section is similar for both reaction, with the ground and $4^+$ excited states taking most of the cross section, although it should be noted that the contribution of the ground state is significantly larger for the $(p,3p)$ reaction than for the two-proton knockout one. To further explore this discrepancy we have performed a calculation in which the interference between configurations with different TNA has been turned off in order to evaluate the effect of this interference in the cross section. The results are shown on the third column of Table.~\ref{tab:28Mgxs}, while the fourth column shows the ratio between these calculations and the original ones. It is noteworthy that it is the ground state the one that presents most sensitivity to this interference, suggesting that it lies behind the difference between $(p,3p)$ and knockout, possibly because $(p,3p)$ is more sensitive to the nuclear interior, which presents a different behaviour than the nuclear surface in respect to the interference of different configurations. This suggests that interference between configurations and correlations between the removed protons plays a significant role in two-proton-removal reactions \cite{Sim10,Sim11}, and that $(p,3p)$ and two-proton knockout reactions may be differently sensitive to them. The last two columns of Table~\ref{tab:28Mgxs} show the sum of the square of the TNA and their ratio to the total for each $^{26}\mathrm{Ne}$ final state, to check whether a direct relation between the reaction cross section and the structure observables, akin to spectroscopic factors for one-nucleon removal reactions, can be established for two-nucleon removal ones. As expected, this is not the case, as the contribution of each final state to the total cross section difers significantly from its ``weight'', measured as the sum of the square of the TNAs. Therefore the reaction theory is essential to be able to predict the relative population of each final state, although a rough relation can be established between TNA and cross sections, in that states with a small sum of $TNA^2$ can be expected to be weakly-populated and a state with a large sum is likely to be significantly populated.

\begin{table*}
    \centering
    \begin{tabular}{cc|cc|cc|cc|c|cc}
     $J^\pi_f$&$E_x$ (MeV)& $\sigma (p,3p)$ (mb) & Ratio&$\sigma (p,3p)$ no interf. (mb) &$\dfrac{(p,3p)^\mathrm{no interf.}}{(p,3p)}$ & $\sigma 2p$KO (mb) &Ratio& Exp $2p$KO (mb) & $\sum \mathrm{TNA}^2$&Ratio
         \\ \hline
          $0^+$& 0.00& 0.274& 0.48&0.196&0.71& 1.19&0.40 &  0.70(15) & 1.28 & 0.24\\
      $2^+$&2.02& 0.048& 0.08&0.048&1.00 & 0.32 & 0.11 & 0.09(15)  &0.69&0.13 \\
       $4^+$&3.50& 0.181 & 0.32&0.160&0.88  & 1.02& 0.34 &  0.58(9)&2.66&0.49\\
       $2^+$&3.70& 0.063& 0.11 &0.061&0.97& 0.45 & 0.15 & 0.15(9)&0.79&0.15\\
         \hline
         Total&&0.57&&0.46&&2.98&&1.50(10)&5.42&
    \end{tabular}
    \caption{Cross sections for the $^{28}$Mg$(p,3p)^{26}$Ne$(J^\pi_f)$ reaction at 250 MeV/A (3rd column) and the$^9$Be$(^{28}$Mg,$^{26}$Ne$(J^\pi_f))X$ reaction at 82.3 MeV/A (7th column \cite{Sim10}, 9th column \cite{Tos04}), and the ratio of each $^{26}$Ne final state to the total cross section (2nd and 6th column respectively). The 3rd column corresponds to calculations where configurations with different TNA are not allowed to interfere and the 4th column is the ratio between this calculation and the full one. The 10th column corresponds to the sum of the square of the TNA considered for each state, and the 11th column to its ratio to the total adding all final states. }
    \label{tab:28Mgxs}
\end{table*}

\subsection{$^{54}\mathrm{Ca}(p,3p)^{52}\mathrm{Ar}$\label{subsec:caar}}

Although there is a number of medium-mass nuclei for which the $(p,3p)$ reaction has been measured \cite{Fro20}, for these nuclei the nuclear structure calculations are computationally heavy and the experimental data scarce, so there are significant uncertainties in the TNA (admittedly, also in the optical potentials). As such, it is challenging to find neutron-rich nuclei for which to compare experimental cross sections to the results of this work. For the $^{54}\mathrm{Ca}(p,3p)^{52}\mathrm{Ar}$ reaction, we were facilitated unpublished experimental data \cite{HongnaPriv} at 250 MeV/A, which yielded a total cross section of $\sigma=0.047(6)$ mb. For $<^{54}\mathrm{Ca}|^{52}\mathrm{Ar}>$, TNA have been graciously provided by Prof. Y.~Utsuno \cite{UtsunoPriv}, and have been obtained using an extension of the GXPF1Br \cite{Ste13} interaction extended to the sd-pf-sdg space. The ones considered in these calculations are presented in Table~\ref{tab:54CaTNA}. When considering which states to include, a significant strength can be found close to the separation energy of the neutron $S_n$ in $^{52}$Ar. Taking the values from NuDat \cite{nudat}, we find a value for $S_n$ of 3.0(7) MeV, obtained by a systematic fit to the nuclear mass, so there is an ambiguity on whether the population of states with excitation energy $E_x=3.0-3.7$ MeV would lead to a bound residual $^{52}$Ar (thus contributing to the cross section), or not.

\begin{table*}
    \centering
    \begin{tabular}{cc|cccccc|cc}
    $J^\pi_f$&$E_x$ (MeV)& $[1d_{5/2}]^2$& $[1d_{5/2}1d_{3/2}]$ & $[1d_{5/2}2s_{1/2}]$ & $[1d_{3/2}]^2$& $[1d_{3/2}2s_{1/2}]$ &$[2s_{1/2}]^2$& $\sigma_{p3p}$ (mb) & $R_s$\\ \hline
      $0^+$& 0.00& -0.221& ---& --- &-0.839 &  --- & -0.222 & 3.51$\cdot 10^{-2}$ & \\
      $2^+$& 1.719& -0.205&-0.238& -0.277&-1.187 &  0.650&--- & 4.82$\cdot 10^{-2}$ &\\
      $0^+$& 2.356& 0.053& ---& --- &0.277 &  --- & -0.030 & 2.75$\cdot 10^{-3}$ & \\
      $2^+$& 2.364& 0.118& 0.074 &0.192&0.575 & -0.492&--- & 1.54$\cdot 10^{-2}$ \\
      $3^+$& 2.809& ---&0.029 &-0.013&--- & --- & ---& 1.35$\cdot 10^{-5}$&\\
      \hline
      \hline
      $2^+$& 3.012& 0.167& 0.077& 0.052 &1.378&-0.147 &  --- & 3.32$\cdot 10^{-2}$ &  \\
      $3^+$& 3.363&---&-0.077 &-0.013&--- & --- & ---& 7.62$\cdot 10^{-5}$& \\
      $4^+$& 3.464&0.073&0.318 &---&--- & --- & ---& 2.57$\cdot 10^{-3}$& \\
      $4^+$& 3.592&0.0511&0.355& --- &---&--- &  --- &3.00$\cdot 10^{-3}$& \\
      $2^+$& 3.639& -0.017& 0.108 &0.263&-0.678&-0.984&---&1.89$\cdot 10^{-2}$&\\
      $0^+$& 3.670& 0.001&---&---&0.089&---&-0.096&2.97$\cdot 10^{-4}$& \\
      \hline \multicolumn{4}{l}{Total ($E_x^\mathrm{max}=S_n=3.0$ MeV)}&&&&&$1.02\cdot 10^{-1}$&0.46\\
       \multicolumn{6}{l}{Total ($E_x^\mathrm{max}=S_n+\sigma_{S_n}=3.7$ MeV)}&&&$1.59\cdot 10^{-1}$&0.29\\
    \end{tabular}
    \caption{States, TNA and cross sections considered for the $^{54}$Ca$(p,3p)^{52}$Ar reaction. The last column corresponds to the reduction factor $R_s=\sigma_\mathrm{exp}/\sigma_{th}$. States above the double horizontal line lie below the nominal value of $S_n$ for $^{52}$Ar, while those below lie below its value plus one $\sigma$ \cite{nudat}. }
    \label{tab:54CaTNA}
\end{table*}

\begin{figure}[tb]
\begin{center}
 {\centering \resizebox*{\columnwidth}{!}{\includegraphics{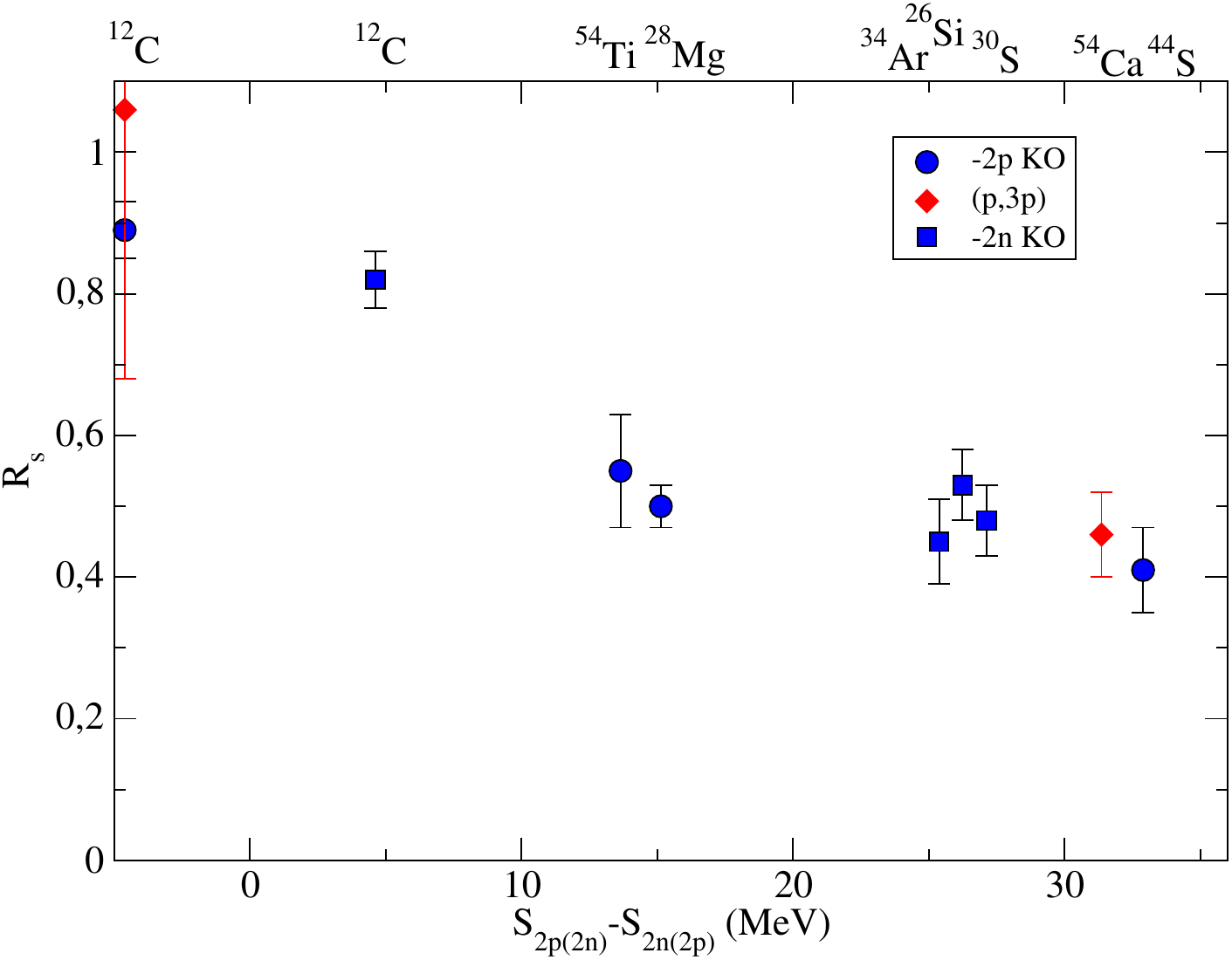}} \par}
\caption{\label{fig:gade}Reduction factors for 2-proton (blue circles) and 2-neutron (blue squares) knockout reactions \cite{Sim10,Tos06,Tos13} and $(p,3p)$ reactions, as a function of $S_{2p}-S_{2n}$ for two-proton removal and  $S_{2s}-S_{2p}$ for two-neutron removal. It can be seen that the two types of reactions follow similar trends.}
\end{center}
\end{figure}

To check the effect of this ambiguity we have computed the cross section to all states for energies up to $3.7$ MeV, and present the cross sections adding up the states up to $E_x=3.0$ MeV (which corresponds to the nominal value of $S_n$) and up to $E_x=3.7$ MeV (which would also include the states within the 1$\sigma$ error range). Both values yield results of $0.102$ mb and $0.159$ mb, which shows that even the uncertainty in $S_n$ produces differences of $50\%$ in the results. Better measurements of the masses of medium-mass nuclei or measurements of $(p,3p)$ reactions with gamma coincidence would help alleviate this ambiguity. As for the reduction factors, we find values of $R_s=0.46$ and $0.29$, both of which are reasonably compatible to those found in \cite{Tos06, Tos13}, even when considering the uncertainties in the separation energy. The reduction factors for two-nucleon knockout \cite{Sim10,Tos06,Tos13} (also including two-neutron knockout from proton-rich nuclei) and $(p,3p)$ reactions are presented in Figure~\ref{fig:gade} as a function of the difference between the separation energy of two protons and two neutrons $\Delta S=S_{2p}-S_{2n}$ for two-proton knockout and $(p,3p)$ and $S_{2n}-S_{2p}$ for two-neutron knockout, as a parallel to the renowned figure for single-nucleon removal \cite{gad08,tos14,Tos21}. $R_s$ for two-proton(neutron) knockout correspond to the blue circles(squares) while the red diamonds correspond to the $(p,3p)$ reactions studied in this work. As can be seen in the figure the tendency is similar for both reactions, showing factors $\sim 1$ for $\Delta S\sim0$ and a significant, roughly constant reduction for more asymmetric nuclei. It is quite remarkable that both reactions show similar trends in the description of the cross sections, while for single-nucleon removal reactions a different trend was found for the reduction factors on $\Delta S=S_{n(p)}-S_{p(n)}$ for one-nucleon knockout and $(p,2p)$ reactions \cite{Aum21}, which has been alleged to originate from deficiencies in the reaction mechanism \cite{Gom23} or in the description of the wavefunctions of the removed nucleons \cite{Ber21,Li22,Ber23}. The fact that the trends reconcile for two-nucleon removal reactions could be related to the stronger peripherality of these reactions \cite{Sim09}, when compared to one-nucleon removal ones, so that the nuclear interior, where the reaction mechanism and the wavefunctions are worse understood, plays a smaller role in the reaction. Further study on these trends is required to clarify these issues.

\section{Summary and conclusions \label{sec:summary}}

In this work, we present the first, to our knowledge, theoretical description of the sequential $(p,3p)$ 2-proton removal direct reaction able to produce cross sections for specific states of the final nucleus based on the two-nucleon amplitudes from nuclear structure calculations. This theory relies on the eikonal description of the interaction between protons and nucleus and the assumption of quasi-free collisions between the incoming proton and the removed protons. We have validated this theory through comparison with experimental data for the $^{12}\mathrm{C}(p,3p)^{10}\mathrm{Be}$ reaction, finding reasonable agreement. We have also studied the $^{28}\mathrm{Mg}(p,3p)^{26}\mathrm{Ne}$ reaction finding parallels with the results found in 2-proton knockout reactions with heavier targets and applied our theory to the $^{54}\mathrm{Ca}(p,3p)^{52}\mathrm{Ar}$ reaction, finding an overestimation of the cross section of a factor $\sim$2-3, which is consistent with results found in 2-proton knockout reactions. This contrasts with the discrepancy found in single-nucleon reactions. Further study on the two-nucleon-removal trends may prove illuminating for the still-unknown causes of the one-nucleon-removal reactions discrepancy.  Application of this formalism to other targets, such as those measured in \cite{Fro20} or \cite{Tan19}, requires the obtainment of two-nucleon amplitudes, which can prove challenging for these heavy nuclei. Further developments of this work may include the use of microscopic optical potentials for the exotic cases where global parametrizations may prove less reliable, as well as the development of a theory to obtain the momentum distribution of the residual target and the angular correlation between the outgoing protons.

\begin{acknowledgements}
 The author thanks Y.~Utsuno for providing the two-nucleon amplitudes used for the $^{54}$Ca case and H.~Liu for the access to the preliminary experimental data for $^{54}$Ca. The author would like to thank as well J.~G\'omez-Camacho, A.M.~Moro and A.~Obertelli for illuminating discussions and critical reading of the manuscript. The author acknowledges financial support by MCIN/AEI/10.13039/501100011033 under I+D+i project No.\ PID2020-114687GB-I00 and under grant IJC2020-043878-I (also funded by ``European Union NextGenerationEU/PRTR''), by the Consejer\'{\i}a de Econom\'{\i}a, Conocimiento, Empresas y Universidad, Junta de Andaluc\'{\i}a (Spain) and ``ERDF-A Way of Making Europe'' under PAIDI 2020 project No.\ P20\_01247, by the European Social Fund and Junta de Andalucía (PAIDI 2020) under grant number DOC-01006 and by the Alexander von Humboldt foundation.
\end{acknowledgements}



\bibliography{p3p.bib}

\end{document}